\documentstyle{laa}

\begin{document}
\thesaurus {08(08.16.5; 08.12.1; 08.12.2);10(10.15.2 Pleiades)}
\title {Brown Dwarfs in the Pleiades Cluster: a CCD-based $R$, $I$ survey
\thanks{Based on observations made with the Jacobus Kapteyn Telescope, 
operated on the island of La Palma by the Royal Greenwich Observatory in 
the Spanish Observatorio del Roque de los Muchachos of the Instituto de 
Astrof\'\i sica de Canarias (IAC); on observations made with the IAC80 
telescope operated by the IAC at its Observatorio del Teide; on 
observations made with the Nordic Optical Telescope in the  Observatorio 
del Roque de los Muchachos; and on observations made with the 2.2~m 
telescope at the Spanish-German Astronomical Center, Calar Alto, 
Almer\'\i a, Spain.}
}

\author {M.R. Zapatero Osorio 
      \and R. Rebolo
      \and E.L. Mart\'\i n}
\offprints {M.R. Zapatero Osorio, e-mail: mosorio@iac.es}
\institute {Instituto de Astrof\'\i sica de Canarias,
            V\'\i a L\'actea s/n, E-38200 La Laguna, Tenerife, Spain}
\date {Received date; accepted date}

\maketitle

\begin{abstract}
We have obtained deep CCD $R$ and $I$ mosaic imaging of 578 arcmin$^{2}$ 
within 1$^{\circ}$.5 of the Pleiades' center -- reaching a completeness 
magnitude $I$~=~19.5 -- with the aim of finding free-floating brown 
dwarfs. Teide~1, the best $bona$ $fide$ brown dwarf discovered so far in 
the cluster (Rebolo, Zapatero Osorio \& Mart\'\i n \cite{rebolo95}), 
arose as a result of a combined photometric and astrometric study of 
$\sim$1/4 of our covered area. The extension of our two-colour survey 
provides eight new additional brown dwarf candidates whose photometry is 
rather similar to that of Teide~1. Several of them are even fainter.  
Follow up low-resolution spectroscopy (Mart\'\i n, Rebolo 
\& Zapatero Osorio \cite{martin96}) shows that one of them is indeed a 
Pleiades brown dwarf. Most of the remaining candidates are background 
late-M dwarfs which are contaminating our survey, possibly due to a 
small (previously unknown) cloud towards the cluster which affects some 
of our CCD fields. We did not expect any foreground M8--M9 field dwarf 
in our surveyed volume and surprisingly we have found one, suggesting 
that its number could be larger than inferred from recent luminosity 
function studies in the solar neighbourhood.


\end{abstract}

\section {Introduction}
Photometric searches for free-floating, very low-mass stars and brown 
dwarfs (hereafter referred to as BDs) in young open clusters have been 
carried out during the last few years (see Jameson \cite{jameson95} 
for a review). The Pleiades has become the favourite target due to its 
scarce reddening and relatively young age compared to other nearby 
clusters. Searches conducted in recently formed clusters take advantage 
of the fact that BDs are still young and bright, and hence, rather easy 
to detect. The expected brightness of all BDs becomes lower as they age. 
For instance, according to recent theoretical evolutionary tracks 
(Burrows et al. \cite{burrows93}; D'Antona \& Mazzitelli 
\cite{dantona94}; Baraffe et al. \cite{baraffe95}), a BD of 
0.06~$M_{\odot}$ at the age of the Pleiades is $\sim$20 times more 
luminous than at the age of the Hyades. Furthermore, the Pleiades 
cluster presents another fortunate circumstance. It is well known that 
the borderline which separates stars from BDs takes place at 
0.08--0.07~$M_{\odot}$ for solar metallicity. In the Pleiades cluster 
this particular mass range coincides with the preservation of Li 
(Magazz\`u, Mart\'\i n \& Rebolo \cite{magazzu93}; Nelson, Rappaport 
\& Chiang \cite{nelson93}; D'Antona \& Mazzitelli \cite{dantona94};) 
which provides a spectroscopic tool for discriminating BDs from stars 
(Rebolo, Mart\'\i n \& Magazz\`u \cite{rebolo92}).

Li abundances in K-type, early and mid M-type stars of the Pleiades 
(Garc\'\i a L\'opez, Rebolo \& Mart\'\i n \cite{garcia94}; Mart\'\i n, 
Rebolo \& Magazz\`u \cite{martin94}; Marcy, Basri \& Graham 
\cite{marcy94}) show that a large depletion has already occurred at the 
age of the cluster. However, one object named PPl~15 (M6.5) discovered by 
Stauffer, Hamilton \& Probst  (\cite{stauffer94}), presents a weak Li 
line suggesting that it has retained some of its initial content (Basri, 
Marcy \& Graham \cite{basri96}). The mass of PPl~15 should be around 
0.08~$M_{\odot}$, which locates this object at the substellar limit in 
the Pleiades.Very recently, another object named Teide Pleiades~1 
(hereafter, Teide~1)was found to have photometry, spectral type, radial 
velocity and proper motion that qualify it as a Pleiades member (Rebolo, 
Zapatero Osorio \& Mart\'\i n \cite{rebolo95}). This object is less 
luminous, cooler and hasa higher Li abundance than PPl~15  (Rebolo et 
al. \cite{rebolo96}), and therefore it is well located in the BD domain. 

Teide~1 was discovered in the early stages of the $R$, $I$ survey that we
present in this paper. The survey has been extended to an area
$\sim$4 times larger than the one covered at that time attaining similar
limiting magnitudes. Eight new objects have been identified with $R$, $I$ 
photometry very similar to that of Teide~1. In Sect.~2, we give
a detailed description of the observations; Sections~3 and 4
present the photometric and proper motion measurements and a discussion
of the results. Finally, conclusions are summarized in Sect.~5.

\section {Observations}
Our $R$, $I$ photometric survey has been carried out using the following 
telescopes: the 0.8~m IAC80 at the Observatorio del Teide; the 1~m 
Jacobus Kapteyn Telescope (JKT) and the 2.5~m Nordic Optical Telescope 
(NOT), both at the Observatorio del Roque de los Muchachos, and the 2.2~m 
at Calar Alto Observatory. The CCDs used were a Thomson 1024$\times$1024 
(IAC80 and NOT) and a Tektronix 1024$\times$1024 (JKT and Calar Alto 
2.2~m), which provided fields of view of 54.5, 5.5, 31.5, and 
22.5~arcmin$^{2}$, respectively. A total of $\sim$578 arcmin$^{2}$ 
($\sim$1\% of the cluster total area) was surveyed in the $R$, $I$ 
broad-band filters. Two exposures of typically 1800~s (IAC80, JKT) and 
300~s (NOT and Calar Alto 2.2~m) were obtained in each filter in order 
to achieve similar limiting magnitudes in all four telescopes. Weather 
conditions during all the nights were always fairly photometric, while 
the seeing ranged between 0''.6 and 2''. In Table~1 we list the log of 
observations as well as the area surveyed by each telescope. 

\begin{table}
\caption[]{Log of photometric observations}
\begin{center}
\begin{tabular}{ccccc}
\hline
Telescope  &      Obs. date & Surv. area   & $R_{\rm lim}$ & $I_{\rm lim}$ \\
           &                &(arcmin$^{2}$)&               &               \\
\hline
IAC80      & 5--7 Jan 1994  & 125          & 22.0          & 21.0\\
NOT        &   7 Nov 1994   &  40          & 22.5          & 21.5\\
C.A. 2.2~m & 26,27 Nov 1994 & 350          & 22.5          & 21.5\\
JKT        & 22,24 Nov 1995 &  63          & 22.5          & 21.5\\
\hline
\end{tabular}
\end{center}
\end{table}

We adopted the $R$, $I$ broad-band filters because BDs were expected to 
be very red objects, ($R-I$) $\geq$ 2.0, that should follow the sequence 
in the $I$ versus $(R-I)$ diagram defined by the mid-M Pleiades stars 
(Hambly, Hawkins \& Jameson \cite{hambly93}). On the other hand, current 
CCD detectors have a high efficiency at $R$ wavelengths and, although it 
drops considerably in the $I$-band, this effect is compensated by the 
exceedingly intense brightness of BDs at these near-IR wavelengths.

We re-observed the fields of Jameson \& Skillen (\cite{jameson89}) using 
the IAC80; therefore, the BD candidates proposed by these authors are 
included in our survey. The goal was to derive proper motions by 
comparing the two epochs of observations separated in time by 
7.17~yr. This is feasible because the cluster members have a large
peculiar proper motion in comparison with field stars in the same region 
of the sky, and hence, proper motion measurements can be achieved within 
a few years. New fields observed with the NOT, JKT and Calar Alto 2.2~m 
telescopes were selected to be roughly 0$^{\circ}$.5--1$^{\circ}$.5 
southeast far from the accepted center of the cluster. New photometric 
candidates were expected to arise from those images. 

According to several recent reddening maps of the Pleiades (van 
Leeuwen \cite{leeuwen83}; Breger \cite{breger87}; Stauffer \& Hartmann 
\cite{stauffer87}), only a fairly small portion of the cluster southwest 
of the cluster center suffers from high absorption. None of our fields 
fall within this region. Therefore, the expected reddening for our BD 
candidates is A$_{I}$~=~0.07~mag and $E$($R-I$)~=~0.03~mag. In Fig.~1 
the location of all the frames imaged during the four observing runs is 
presented to scale. Stars brighter than 6$^{th}$~magnitude and M stars 
(Hambly et al. \cite{hambly93}) within 2$^{\circ}.5\times$2$^{\circ}$.5 
centered at $\sim$3$^{\rm h}47^{\rm m}$, 24$^{\circ}$7' (Eq. 2000) are 
also included for comparison. Although the region covered by our survey 
only represents a small fraction of the total area of the cluster, 
several interesting objects have been discovered (see the next sections). 

\begin{figure}
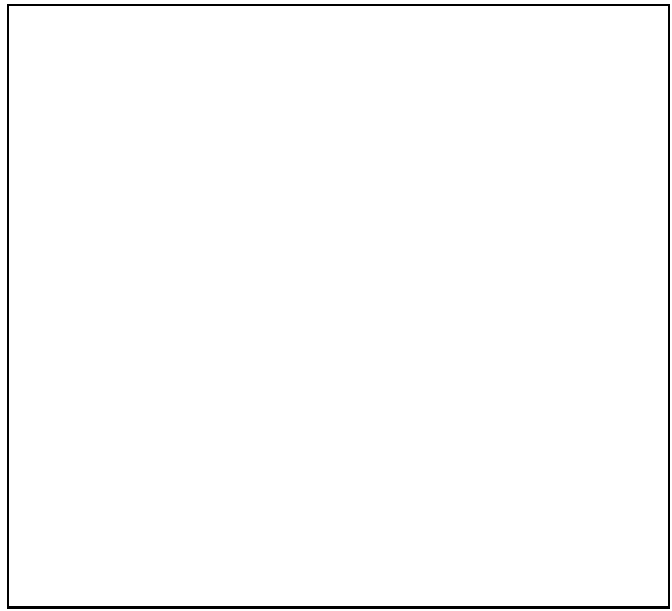

\picplace{8 cm}
\caption[]{Location of our fields (open squares) within 
2$^{\circ}.5\times$2$^{\circ}$.5 of the Pleiades area. Central 
coordinates are $\sim$3$^{\rm h}47^{\rm m}$, 24$^{\circ}$7' (Eq. 2000). 
Filled circles stand for stars brighter than 6$^{th}$~ magnitude and 
for proper motion M members (Hambly et al. \cite{hambly93}) with $I$ 
magnitudes in the range 13--17. The vertical gap in the M~star 
distribution around 3$^{\rm h}51^{\rm m}$ is due to the fact that there 
were no overlaps between the first and second epoch plates used by the 
authors, causing the lack of proper motion measurements for stars in 
that strip. The relative brightness is represented by circle diameters. 
North is up and East is left}
\end{figure}

\section{Data analysis}
\subsection{Photometry}
Raw frames were processed using standard techniques within the IRAF 
\footnote{IRAF is distributed by National Optical Astronomy Observatories,
which is operated by the Association of Universities for Research in
Astronomy, Inc., under contract with the National Science Foundation.}
environment, which included bias substraction, flat-fielding and 
correction for bad pixels by interpolation with values from the 
nearest-neighbour pixels. The photometric analysis was carried out 
using routines within DAOPHOT, which provides image profile information 
needed to discriminate between stars and galaxies. Instrumental 
magnitudes were corrected for atmospheric extinction and transformed 
into the $R$, $I$ Cousins (\cite{cousins76}) system using observations 
of standard stars from Landolt's (\cite{landolt92}) list. Fields taken 
with the NOT and Calar Alto 2.2~m telescopes were overlapped by 0'.5 
and 1'.5 respectively. On the basis of the photometry of stars that 
fall in overlapping regions of adjacent frames, we estimate the 
uncertainties to range from $<$0.05 mag at $R$, $I$ $\sim$ 18.5 to 
about 0.15~mag at 21.0~mag. Because no standard star beyond 
$(R-I)$~=~1.6 was included in the transformation equations, the 
uncertainties for the reddest, faintest objects are expected to be 
slightly larger. 

Limiting magnitudes (listed in Table~1) for all the telescopes are quite 
alike as exposure times were scaled according to the telescope diameters. 
However, different seeing conditions caused the completeness magnitudes 
to differ not only from telescope to telescope but also from one 
exposure to another. NOT frames were always obtained in good seeing 
($<$1'') and therefore, go deeper than the IAC80, JKT and Calar Alto 
2.2~m images. The comparison of the total number of stars per magnitude 
interval in each telescope to that observed with the NOT provides us 
with an accurate determination of the completeness of our survey. We 
estimate it to be $R$, $I$~=~20.0, 19.0 (IAC80) and $R$, $I$~=~20.5, 
19.5 (JKT, NOT and Calar Alto 2.2~m). These results were confirmed by 
the comparison with predictions of the number of stars present in the 
Galaxy by Bahcall \& Soneira (\cite{bahcall81}) and by Gilmore \& Reid 
(\cite{gilmore83}).

\begin{figure}
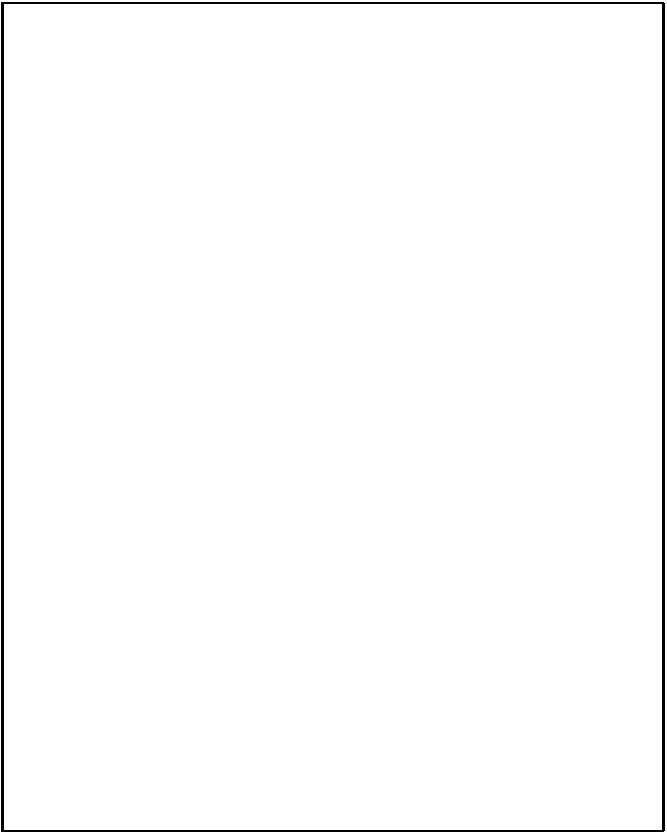

\picplace{11 cm}
\caption[]{Colour-magnitude diagram for very low-mass stars and BDs in 
the Pleiades cluster. Dots stand for proper motion M Pleiads from Hambly 
et al. (\cite{hambly93}); filled squares stand for the nine BD 
candidates of Jameson \& Skillen (\cite{jameson89}); open and filled 
circles stand for objects in our survey. The location of the substellar 
limit is indicated by PPl~15. The faintest BD in the cluster with proper 
motion measured, Teide~1, is also plotted} 
\end{figure}

We present in Fig.~2 the $I$ vs ($R-I$) diagram for the Pleiades JKT, 
NOT and Calar Alto 2.2~m fields, combining data from Hambly et al. 
(\cite{hambly93}) with the new observations. These authors published 
photographic magnitudes which were converted into the Cousins system 
using the transformation equations given in Bessell (\cite{bessell86}). 
The straight line that separates Pleiads from field stars was derived by 
taking into account those previously known proper motion members. No 
reddening has been applied. To indicate the position of the 
stellar-substellar borderline in the Pleiades, we have included PPl~15 
for which $R$~=~20.05$\pm$0.07 was measured during the observing run 
carried out with the JKT. Its $I$ magnitude was taken from Stauffer 
et al. (\cite{stauffer94}). Also shown in Fig.~2 is the BD Teide~1 (see 
next subsection). Because its substellar nature has been confirmed 
recently (Rebolo et al. \cite{rebolo96}), objects found to present 
similar photometry to that of Teide~1 and located above the straight 
line might be BDs as well. 

\begin{table}
\caption[]{Brown dwarf candidates in the Pleiades}
\begin{center}
\begin{tabular}{lccclc}
\hline
\multicolumn{1}{c}{Name} &
\multicolumn{2}{c}{RA (J2000) DEC} &
\multicolumn{1}{c}{$I$}  &
\multicolumn{1}{l}{$R$--$I$}\\
\multicolumn{1}{c}{} &
\multicolumn{1}{c}{($^{\rm h}$  $^{\rm m}$ $^{\rm s}$)} &
\multicolumn{1}{c}{($^{\circ}$ ' \ '')} &
\multicolumn{1}{c}{} &
\multicolumn{1}{l}{} \\
\hline
Calar Pleiades 1          & 3 51 04.0 & 23 51 02 & 18.18 & 2.93   \\
Roque Pleiades 1          & 3 50 00.0 & 23 34 03 & 18.43 & 2.50   \\
Calar Pleiades 2          & 3 51 15.0 & 23 54 01 & 18.66 & 2.50   \\
Calar Pleiades 3          & 3 51 26.0 & 23 45 20 & 18.73 & 2.54:  \\
Teide Pleiades 1$^{\ast}$ & 3 47 18.0 & 24 22 31 & 18.80 & 2.74   \\
Calar Pleiades 4          & 3 50 43.0 & 23 52 46 & 18.88 & 2.37   \\
Calar Pleiades 5          & 3 50 52.1 & 23 51 48 & 19.01 & 2.33   \\
Calar Pleiades 6          & 3 51 21.0 & 23 52 10 & 19.20 & 2.59:  \\
Calar Pleiades 7          & 3 51 21.3 & 23 52 10 & 19.7: & 2.52:: \\
\hline
\end{tabular}
\end{center}
{\bf Notes.} \\
$^{\ast}$~Teide Pleiades~1 is a proper motion member.\\
For uncertainties in photometry and coordinates, see text.
Colours labelled with a colon have uncertainties of $\pm$0.20 mag;
the one labelled with two colons, $\pm$0.35 mag.\\
\end{table}

Table~2 provides the names, magnitudes, colours and positions for the new 
proposed Pleiades BD candidates. They are named according to the 
observatory in which they were first detected followed by the word 
$Pleiades$, and numbered according to their increasing $I$-band apparent 
magnitude. Hereafter, we will use an abridged version of the names which 
omits the term $Pleiades$. Coordinates are accurate to approximately 
$\pm$2''.5. Finding charts (2'$\times$2' in extent, $I$-band) are 
provided in Fig.~3.

\begin{figure*}
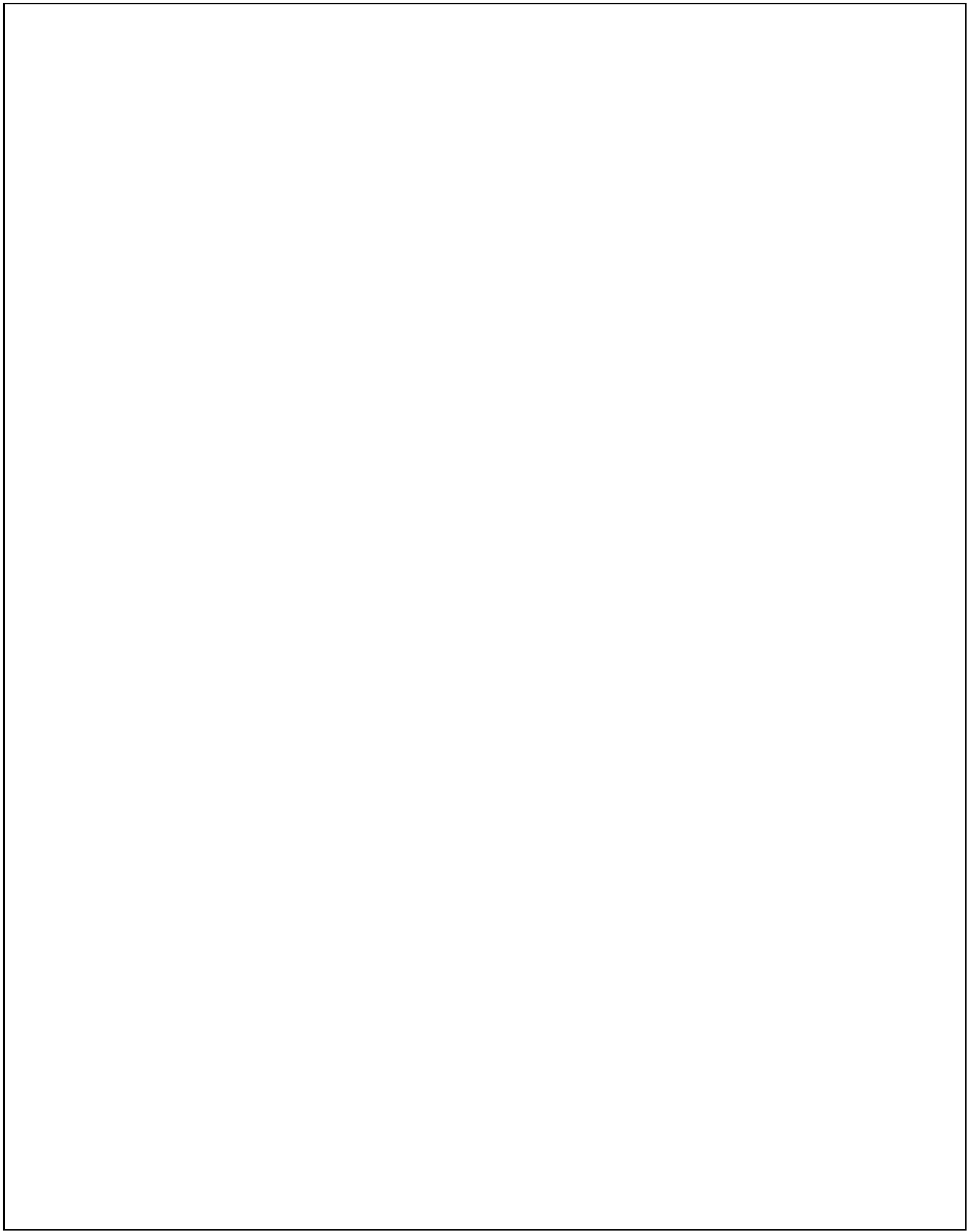

\picplace{23 cm}
\caption[]{Finding charts (2'$\times$2' in extent, $I$-band) for the BD 
candidates listed in Table~2 and Teide~1. North is up and East is left}
\end{figure*}

\subsection{Proper motion survey}
Approximately 25\% of the total area covered in our survey was 
chosen to overlap regions previously observed by Jameson \& Skillen 
(\cite{jameson89}). We were looking for BD candidates whose photometry 
and proper motion support membership in the Pleiades cluster. For 
proper motion measurements we have mainly used $I$-band frames because 
the objects which interest us are brighter in this filter than in $R$ 
and, therefore, their centroids can be determined more accurately. The 
astrometric procedures for deriving proper motions were those described 
by Hawkins (\cite{hawkins86}) and have been used successfully by Hambly 
et al. (\cite{hambly93}, \cite{hambly95}). We estimate that our survey 
is complete in the range 16.5 $<$ $I$ $<$ 19~mag. 

After having covered 125~arcmin$^{2}$ around the center of the cluster, 
a sole object arose with photometric and proper motion determinations 
perfectly compatible within the error bars with those that identify the 
cluster. It was only detected in the $I$-band ($I$~=~18.80$\pm$0.07) 
suggesting that its colour should be very red [($R$--$I$) $>$ 2.2]. We 
named it as Teide~1 (Rebolo et al. \cite{rebolo95}). The field in which 
Teide~1 appeared was re-observed again during the NOT campaign (8 years 
after Jameson \& Skillen's run), yielding ($R$--$I$)~=~2.74$\pm$0.10~mag. 
The $R$, $I$ photometry and coordinates for Teide~1 are listed in 
Table~2. From November 1986 to November 1994, the measured proper motion 
of Teide~1 was $\mu_{\alpha {\rm cos} \delta}$~=~0.13''$\pm$0.07'' and 
$\mu_{\delta}$~=~--0.28''$\pm$0.13'', which is consistent within the 
error bars with that of the Pleiades cluster for the same time interval 
(that is $\mu_{\alpha {\rm cos} \delta}$~=~0.20''$\pm$0.05'' and 
$\mu_{\delta}$~=~--0.36''$\pm$0.05'', Jones \cite{jones81}). Teide~1 
has been studied spectroscopically by Rebolo et al. (\cite{rebolo95}), 
Rebolo et al. (\cite{rebolo96}) and Mart\'\i n, Rebolo \& Zapatero Osorio 
(\cite{martin96}). They conclude that the measured proper motion and 
radial velocity, the H$\alpha$ emission, the M8--M9 spectral type, the 
$R$, $I$ photometry as well as the fact that it has preserved a large 
amount of Li in its atmosphere, are all consistent with Teide~1 being 
a Pleiades BD. 

The nine CCD survey objects of Jameson \& Skillen (\cite{jameson89}) were 
present in our proper motion analysis. Bearing in mind that PPl~15 fixes 
the frontier of the very low-mass stars in the cluster, only JS1 
(=~PPl~3, Stauffer et al. \cite{stauffer89}) and JS2 might be BD 
candidates according to their $I$ magnitude. Our photometric analysis 
for JS1--9 yielded within the error bars similar magnitudes and colours 
than those previously published. Nevertheless, they (except JS9) lie 
about 0.5--1~mag below the straight line in Fig.~2 (i.e. 1.5--2~mag below 
the Pleiades sequence), suggesting that these objects are probable 
non-members according to their $R$, $I$ photometry. 
From previous works, it is known that JS4, 6, 8 and 9 failed as proper 
motion members (Hambly et al. \cite{hambly91}). JS1, 2, 3, 5 and 7 were 
either too faint or blended to be included in their study and they were 
hence rejected. JS1 is unlikely to be a member due to its IR photometry 
(Stauffer et al. \cite{stauffer89}; Hambly et al. \cite{hambly91}). 
Stringfellow (\cite{stringfellow91}) argues that none of them are members 
on the basis of their location on the theoretical HR diagram. Our proper 
motion study does not provide support for their membership regardless of 
their photometry.

\section{Discussion}

With $R$, $I$ photometric data only we cannot conclude that our BD 
candidates listed in Table 2 are true Pleiads. We will discuss briefly 
what kind of objects are expected to be contaminating a survey like ours 
and will try to quantify each of them as follows. 

Distant galaxies may be one source of contamination. However, we
feel confident that our new BD candidates are stellar-like objects 
because most galaxies with $I$ magnitudes around $I$~=~18--19 might be 
resolved in our survey given the pixel size of the detectors and seeing 
conditions. Therefore, they would not have fitted the point spread 
function, and our BD candidates certainly did. The recent $I$, $K$ survey 
by Jameson et al. (\cite{jameson96}) has shown that the primary 
contaminants are faint, red galaxies. It seems that the use of optical 
and near-IR filters allow to better discriminate BDs from these 
presumably distant galaxies, despite the fact that BDs are much fainter 
at these wavelengths than at $K$-band. 

Another source of contamination comes from background late-M giants. 
We can estimate this kind of contribution given the recent study of the 
luminosity function for free-floating M stars at the end of the main 
sequence by Kirkpatrick et al. (\cite{kirkpatrick94}). From their 
survey (limiting magnitude $R$~=~19), the authors argue that these 
objects are only relatively numerous in areas within 10$^{\circ}$ of 
the Galactic plane. As the Pleiades cluster is located well away from 
this region (at $b$~=~--24$^{\circ}$), the expected number of background 
late-M giants in our survey is negligible. 

Due to their location in the $I$ vs ($R-I$) diagram, our BD candidates 
could be Pleiades substellar objects, or old field M7.5--M9 dwarfs just 
superimposed on the cluster or else background, reddened field M5--M7 
dwarfs. We have used the luminosity function for main sequence 
stars in the solar neighbourhood of Kroupa (\cite{kroupa95}) in order 
to estimate the number of M5--M7 dwarfs that could be present in the 
volume that our survey has covered. For star counts out of the plane of 
the Galaxy the space density is well represented by an exponential law. 
Adopting the scale height for dwarf M stars given by Mihalas \& Binney 
(\cite{mihalas81}), a total of seven stars may be contaminating, one of 
which could be a foreground star. Furthermore, M5--M7 dwarfs were found 
to be rather abundant in Kirkpatrick et al.'s (\cite{kirkpatrick94}) 
work: 33 dwarfs out of 95 turned out to have spectral types within this 
range (i.e. 35\%). On the contrary, only one M8--M9 star was discovered, 
suggesting that the local density for this spectral range is 
$\sim$0.0024~pc$^{-3}$. Because our survey in the JKT, NOT and Calar 
Alto 2.2~m covered a volume about 10 times smaller, the probability of 
finding such late spectral type objects along the line of sight towards 
the cluster is almost negligible. We have obtained similar results using 
the luminosity functions of different authors (Leggett \& Hawkins 
\cite{leggett88}; Tinney \cite{tinney93}). Therefore, based on these 
statistics the main source of contaminating objects in our survey 
comes from M5--M7 dwarfs. 

Mart\'\i n et al. (\cite{martin96}) have conducted low-resolution 
spectroscopic observations of all our BD candidates. The likelihood of 
membership was assessed via the study of their spectroscopic properties 
compared to some very late field M stars and Teide~1. As expected from 
previous estimates, all the BD candidates are late-M dwarfs. However, 
only Calar~3 turns out to be an object very similar to Teide~1 and 
therefore, a very likely Pleiades BD. Based on the available data 
this object clearly meets all the membership criteria. The H$\alpha$ 
emission, radial velocity, spectral type and photometry are consistent 
with Calar~3 being a member of the cluster. This result yields a success 
rate of two genuine BDs out of our nine candidates ($\sim$22.5\%). If 
an uniform distribution within the cluster is assumed, the expected 
number of Pleiades BDs with $I$~=~18.5--19.0 would be about 200 objects, 
suggesting that the probable total number of BDs in the substellar 
domain is rather large. The implications of this result for the mass 
function are beyond the scope of this paper and will be addressed in a 
future work which takes into account not only our survey, but all the 
surveys carried out so far in the cluster.  

Other very interesting objects have arisen as by-products of our survey. 
Calar~1 and Roque~1 are surprisingly high radial velocity, 
very late-M dwarfs and hence, probably non-members that need further 
observations in order to understand their nature. To our knowledge, 
Calar~1 (M9) and Roque~1 (M7) have the highest radial velocity ever 
measured among the latest dwarfs. According to their spectral types and 
$I$ apparent magnitudes, Calar~1 and Roque~1 should be located at 
approximately 45~pc and 111~pc respectively, i.e. foreground objects. 
The existence of an M7 dwarf at this distance in our survey was somehow 
expected based on previous surveys, whereas the discovery of an M9 dwarf 
remarkably disagrees with expectations. Its detection could be indicative 
of a large, yet undiscovered, population of field M9 dwarfs in the solar 
neighbourhood. 

Calar~2,~4,~5,~6 and~7 are likely M4--M6.5 background, reddened dwarfs 
rather than members of the cluster. These stars are the main source of 
contamination in our colour-magnitude diagram. All of them lie within a 
quite small region of the sky ($\sim$20~arcmin$^{2}$) as shown in Fig.~4, 
suggesting that this area may suffer from enhanced reddening. We have 
estimated the mean colour excess, $E(R-I)$, in this little region by 
comparison of the observed ($R-I$) colour with the reddening-free colour 
for each spectral type given by Kirkpatrick \& McCarthy 
(\cite{kirkpatrick94}). Calar~7 was excluded from the calculations due to 
the large uncertainty in its photometry. The mean reddening is determined 
to be $E(R-I)$~=~0.30$\pm$0.09~mag, possibly indicating the existence of 
a cloud towards the Pleiades at $\alpha$~=~3$^{\rm h}$51$^{\rm m}$, 
$\delta$~=~23$^{\circ}$53' (Eq.~2000). Whether this cloud is either 
foreground, or background or within the cluster still remains unknown. 
We note that the BD Calar~3, located at $\sim$9' distance, is unlikely to 
be affected by such a high reddening since its spectra and photometry are 
extremely similar to those of Teide~1.

\begin{figure}
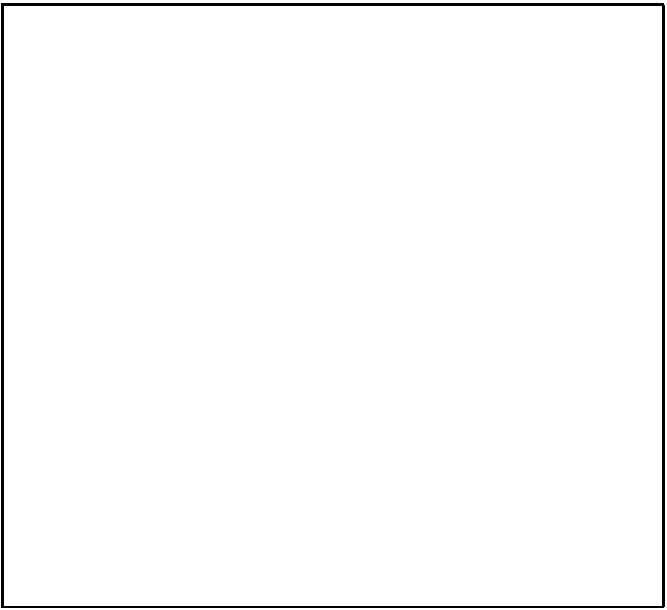

\picplace{8 cm}
\caption[]{Spatial distribution of our Calar BD candidates (filled dots) 
and the fields observed with the 2.2~m Calar Alto telescope (scaled open 
squares). Central coordinates are those of Calar~1 (the plot scale is 
25''/mm). North is up and East is left}
\end{figure}

According to the degree of contamination by field M dwarfs found in 
our survey, we conclude that determinations of the initial mass function 
based on candidates whose membership relies only on two-colour 
photometry are likely to significantly overestimate the number of very 
low-mass members in the cluster. Further analysis is needed in order to 
determine membership and consequently derive reliable mass 
functions. In Mart\'\i n et al. (\cite{martin96}) we have proved that a 
$R$, $I$ photometric survey followed up by low-resolution spectroscopy 
constitutes a reliable tool to assess membership to the cluster.
Moreover, we will also explore the potential of IR photometry to 
characterize Pleiades BDs (Zapatero Osorio, Rebolo \& Mart\'\i n 
\cite{zapatero96}). 

Calar~3 and Teide~1 are young BDs in the Pleiades cluster with quite a 
high level of confidence. The general shape of their optical and near-IR 
spectra is very similar to that of field M8--M9 dwarfs (see Rebolo et al. 
\cite{rebolo95}; Mart\'\i n et al. \cite{martin96}). However, we have 
found that the ($R-I$) colour of Calar~3 and Teide~1 is 0.2--0.4~mag 
redder than those of field dwarfs with the same spectral type. A more 
depressed pseudocontinuum at the $R$-band (fainter $R$'s) is observed 
in the low-resolution spectra of Calar~3 and Teide~1, suggesting that 
the colour offset is real. We have computed the ($R-I$) differences 
between Calar~3, Teide~1 and vB10, LP412--31 LHS2065 using the published 
spectra (Rebolo et al. \cite{rebolo95}; Mart\'\i n et al. 
\cite{martin96}) and found them to be in good agreement with the 
measured photometric differences within the uncertainty bars. This 
suggests that although systematic errors in the photometry could exist 
since our BDs are much redder than the reddest standard star used for 
the calibrations, they should be small. Possible explanations that 
could account for the colour offset may be based on the lower gravity 
in the Pleiades' members than in old dwarfs, and on its relationship 
with the dust formation in the atmospheres of these very cool objects 
(Tsuji, Ohnaka \& Aoki \cite{tsuji96}).

\section{Conclusions}
We have conducted $R$, $I$ observations of 578~arcmin$^{2}$ in the 
Pleiades ($\sim$1\% \ of the cluster area) reaching completeness 
magnitudes $R$, $I$~=~20.5, 19.5~mag and limiting magnitudes $\sim$1~mag 
fainter. As a result of a combined photometric and proper motion survey 
covering 125~arcmin$^{2}$ around the cluster's center (i.e. $\sim$1/4 
of the total surveyed area), we found Teide~1 (Rebolo et al. 
\cite{rebolo95}), whose substellar nature has been confirmed recently by 
the Li test (Rebolo et al. \cite{rebolo96} ). The extension of our 
two-colour survey yielded eight new BD candidates which lie along the 
substellar sequence of the Pleiades and whose $R$, $I$ photometry is 
rather similar to that of Teide~1. 

Further spectroscopic studies 
(Mart\'\i n et al. \cite{martin96}) of our BD candidates reveal that 
$\sim$25\% \ are true BDs (Teide~1 and Calar~3) showing that our 
technique is successful in finding substellar objects. The remaining 
Calar objects and Roque~1 are likely to be late-M field dwarfs. 
Particularly, the fact that a field M9 dwarf (Calar~1) was found in our 
survey may be indicative of a large, yet undiscovered, population of 
these very late and cool objects in the solar neighbourhood. We have 
found that Calar~3 and Teide~1 (spectral type M8), the two genuine BDs 
in the Pleiades, are redder in the ($R-I$) colour than field stars of 
the same spectral type in the sense that they have fainter 
$R$-magnitudes. We argue that this feature could be an effect of the 
lower gravity of Pleiades BDs because of the  youth of these objects.

\acknowledgements
{It is a pleasure to acknowledge N. Hambly for his help with the proper 
motion determination. We thank M. Murphy for her careful reading of the 
manuscript and English corrections. We also thank J. Stauffer 
(referee) for his valuable comments. Partial financial support was 
provided by the Spanish DGICYT project no. PB92--0434--C02.}


\begin{thebibliography}{}     
\bibitem[1981]{bahcall81}
Bahcall, J., Soneira, R., 1981, ApJS, 47, 357
\bibitem[1995]{baraffe95}
Baraffe, I., Chabrier, G., Allard, F., Hauschildt, P.H., 1995, ApJ, 446, L35
\bibitem[1996]{basri96}
Basri, G., Marcy, G.W., Graham, J.R., 1996, ApJ, 458, 600
\bibitem[1986]{bessell86}
Bessell, M.S., 1986, PASP, 98, 1303
\bibitem[1987]{breger87}
Breger, M., 1987, ApJ, 319, 754
\bibitem[1993]{burrows93}
Burrows, A., Hubbard, W.B., Saumon, D., Lunine, J.I., 1993, ApJ, 406, 158
\bibitem[1976]{cousins76}
Cousins,, A.W.J., 1976, Mem. R. Astron. Soc., 81, 25
\bibitem[1994]{dantona94}
D'Antona, F., Mazzitelli, I., 1994, ApJS, 90, 467
\bibitem[1994]{garcia94}
Garc\'\i a L\'opez, R.J., Rebolo, R., Mart\'\i n, E.L., 1994, A\&A, 282, 518
\bibitem[1983]{gilmore83}
Gilmore, G., Reid, N., 1983, MNRAS, 202, 1025
\bibitem[1991]{hambly91}
Hambly, N.C., Hawkins, M.R.S., Jameson, R.F., 1991, MNRAS, 253, 1
\bibitem[1993]{hambly93}
Hambly, N.C., Hawkins, M.R.S., Jameson, R.F., 1993, A\&AS, 100, 607
\bibitem[1995]{hambly95}
Hambly, N.C., Steele, I.A., Hawkins, M.R.S., Jameson, R.F., 1995, A\&AS, 
  109, 1
\bibitem[1986]{hawkins86}
Hawkins, M.R.S., 1986, MNRAS, 223, 845
\bibitem[1996]{jameson96}
Jameson, R.F., Hodgkin, S.T., Pinfield, D., 1996, in the 9th Workshop on 
  Cool Stars, Stellar Systems and the Sun. Florence, 3--6 October 1995, 
  in press
\bibitem[1995]{jameson95}
Jameson, R.F., 1995, in The Bottom of the Main Sequence and Beyond, 
  ESO Workshop, ED. Tinney, p. 183
\bibitem[1989]{jameson89}
Jameson, R.F., Skillen, I., 1989, MNRAS, 239, 247
\bibitem[1981]{jones81}
Jones, B.F., 1981, AJ, 86, 290
\bibitem[1994]{kirkpatrick94}
Kirkpatrick, J.D., McGraw, J.T., Hess, T.R., Liebert, J., McCarthy, D.W., 
  1994, ApJS, 94, 749
\bibitem[1994]{kirkpatrick94}
Kirkpatrick, J.D., McCarthy, D.W., 1994, AJ, 107, 333
\bibitem[1995]{kroupa95}
Kroupa, P., 1995, ApJ, 453, 350
\bibitem[1992]{landolt92}
Landolt, A.U., 1992, AJ, 104, 340
\bibitem[1988]{leggett88}
Leggett, S.K., Hawkins, M.R.S., 1988, MNRAS, 234, 1065
\bibitem[1993]{magazzu93}
Magazz\`u, A., Mart\'\i n, E.L., Rebolo, R., 1993, ApJ, 404, L17
\bibitem[1994]{marcy94}
Marcy, G.W., Basri, G., Graham, J.R., 1994, ApJ, 428, L57
\bibitem[1994]{martin94}
Mart\'\i n, E.L., Rebolo, R., Magazz\`u, A., 1994, ApJ, 436, 262
\bibitem[1996]{martin96}
Mart\'\i n, E.L., Rebolo, R., Zapatero Osorio, M.R., 1996, ApJ, in press
\bibitem[1981]{mihalas81}
Mihalas, D., Binney, J., 1981, in Galactic Astronomy: Structure and 
  Kinematics (2d ed.; San Francisco: W.H. Freeman), 252
\bibitem[1993]{nelson93}
Nelson, L.A., Rappaport, S., Chiang, E., 1993, ApJ, 413, 364
\bibitem[1992]{rebolo92}
Rebolo, R., Mart\'\i n, E.L., Magazz\`u, A., 1992, ApJ, 389, L83
\bibitem[1995]{rebolo95}
Rebolo, R., Zapatero Osorio, M.R., Mart\'\i n, E.L., 1995, Nat, 377, 129
\bibitem[1996]{rebolo96}
Rebolo, R., et al., 1996, in preparation
\bibitem[1994]{stauffer94}
Stauffer, J.R., Hamilton, D., Probst, R .G., 1994, AJ, 108, 155
\bibitem[1989]{stauffer89}
Stauffer, J.R., Hamilton, D., Probst, R .G., Rieke, G., Mateo, M., 1989, 
  ApJ, 344, L21
\bibitem[1987]{stauffer87}
Stauffer, J.R., Hartmann, L., 1987, ApJ, 318, 337
\bibitem[1991]{stringfellow91}
Stringfellow, G.S., 1991, ApJ, 375, L21
\bibitem[1993]{tinney93}
Tinney, C.G., 1993, ApJ, 414, 279
\bibitem[1996]{tsuji96}
Tsuji, T., Ohnaka, K., Aoki, W., 1996, A\&A, 35, L1
\bibitem[1983]{leeuwen83}
van Leeuwen, F., 1983, Ph.D. Thesis, Univ. Leiden
\bibitem[1996]{zapatero96}
Zapatero Osorio, M.R., Rebolo, R., Mart\'\i n, E.L., 1996, in preparation
\end{thebibliography}
\end{document}